\documentclass[aps,pra,superscriptaddress,amsmath,twocolumn,amssymb]{revtex4}

\usepackage{xr}

\usepackage{subfigure}
\usepackage{amsmath}
\usepackage{pgfplots}
\pgfplotsset{compat=1.5}
\usepgfplotslibrary{groupplots}

\usepackage{color}

\usepackage{xcolor}
\usepackage{bm}
\usepackage{multirow}
\usepackage{placeins}
\usepackage{soul}
\usepackage{color,xcolor}
\usepackage[colorlinks,linkcolor=red,anchorcolor=blue,citecolor=blue,urlcolor=blue]{hyperref}
\usepackage{cleveref}

\newcommand{\ccenter}{\vec{\mathcal{O}}}
\newcommand{\qcenter}[1]{\vert \mathcal{O}_{#1}\rangle}
\newcommand{\mcenter}[1]{\mathcal{O}_{#1}^{\vec{\sigma}}}
\newcommand{\mcenterbra}[1]{\mathcal{O}_{#1}^{\vec{\sigma}, \dagger}}

\newcommand{\circuit}[1]{\mathcal{C}_{#1}(\vec{\theta})}
\newcommand{\argmax}{\operatornamewithlimits{argmax}}
\newcommand{\dQ}{d^Q}
\newcommand{\dMPS}{d^M}
\newcommand{\dC}{d^C}

\begin{document}

\title{Quantum inspired K-means algorithm using matrix product states}

\author{Xiao Shi}
\affiliation{Institute of Mathematics, Academy of Mathematics and Systems Science, Chinese Academy of Sciences, Beijing
100190, China}
\affiliation{School of Mathematical Sciences, University of Chinese Academy of Sciences, Beijing 100049, China}

\author{Yun Shang}
\email{shangyun602@163.com}
\affiliation{Institute of Mathematics, Academy of Mathematics and Systems Science, Chinese Academy of Sciences, Beijing
100190, China}
\affiliation{NCMIS, MDIS, Academy of Mathematics and Systems Science, Chinese Academy of Sciences, Beijing, 100190,China}

\author{Chu Guo}
\email{guochu604b@gmail.com}
\affiliation{Department of Physics and Synergetic Innovation Center for Quantum Effects and Applications, Hunan Normal University, Changsha 410081, China}

\begin{abstract}
Matrix product state has become the algorithm of choice when studying one-dimensional interacting quantum many-body systems, which demonstrates to be able to explore the most relevant portion of the exponentially large quantum Hilbert space and find accurate solutions. Here we propose a quantum inspired K-means clustering algorithm which first maps the classical data into quantum states represented as matrix product states, and then minimize the loss function using the variational matrix product states method in the enlarged space. We demonstrate the performance of this algorithm by applying it to several commonly used machine learning datasets and show that this algorithm could reach higher prediction accuracies and that it is less likely to be trapped in local minima compared to the classical K-means algorithm.
\end{abstract}

\date{\today}
\maketitle

\address{}

\vspace{8mm}

\section{Introduction}
The past few years have witnessed a growing interest in the intersection between quantum physics and machine learning. On the one side, machine learning tools have been used to solve quantum problems, such as phase recognition~\cite{CarrasquillaMelko2017,ZhangZhai2018,ChngKhatami2018, NieuwenburgHuber2017,BroeckerTrebst2017, ChngKhatami2017, ZhangKim2017, DongZhang2019}, quantum state tomography~\cite{Torlai2018, Torlai2018b,Rocchetto2018,Quek2018,Carrasquilla2019}, solving quantum many-body problems~\cite{ArsenaultMillis2014,ArsenaultMillis2015,TorlaiMelko2016,AminMelko2018,LiuFu2017,HuangWang2017,AokiKobayashi2016,CarleoTroyer2017,NomuraImada2017,CzischekGasenzer2018}, non-Markovian quantum dynamics~\cite{LuchnikovFilippov2020}. Connections have been drawn between quantum many-body algorithms and neural network algorithms~\cite{Beny2013,MehtaSchwab2014,LinRolnick2017,DengSarma2017}. On the other side, tools from quantum many-body physics, especially tensor network states algorithms, are used to solve classical machine learning problems. Tensor network states algorithms have been extensively used to study quantum many-body systems in the last two decades. In particular, Matrix Product States (MPS), which is a one-dimensional variant of TNS, has achieved great success and become a standard numerical tool to solve one-dimensional strongly correlated quantum many-body systems due to its high efficiency and accuracy~\cite{Schollwock2011}. Applications of MPS to solve machine learning tasks include classification problems~\cite{StoudenmireSchwab2016,Stoudenmire2018,SunSu2020}, generative modeling~\cite{HanZhang2018}, sequence to sequence modeling~\cite{GuoPoletti2018}, where it is shown that MPS based algorithm could achieve a learning precision close to or even better than state of the art machine learning algorithms. Tensor network states based machine learning algorithms has also been applied for quantum process tomography\cite{GuoKavan2020, TorlaiAolita2020}.

In this work, we apply matrix product states to the clustering task, which is an elementary machine learning task to separate unlabeled data into distinct and non-overlapping clusters. A standard algorithm for clustering is the K-means algorithm, which divides the data into $k$ different classes by minimizing the variance of the data in each class~\cite{MacQueen1967}. The most popular K-means clustering algorithm, Lloyd's algorithm, starts from $k$ random centroids, and then iteratively dividing the data into $k$ classes according to their distances with the centroids, and recomputing the center within each cluster~\cite{Lloyd1982}. This algorithm becomes a standard algorithm for clustering due to its simplicity and efficiency, and we will refer to it as the classical K-means algorithm in the rest of this work. However, it is known that in some cases this algorithm could result in bad clustering with high probability, even for randomized initialization. Various efforts have been paid to improve the accuracy of Lloyd's algorithm, such as a clever way of initialization~\cite{ArthurVassilvitskii2007, Ostrovsky2012, Celebi2013, Bachem2016}, or a smoother objective loss function~\cite{Zhang1999, XuLange2019, DeAmorimMirkin2012,ChakrabortyDas2017}.

Similar to other MPS-based machine learning algorithms, the basic idea of this work is to first map the classical data into quantum states which live in an exponentially large Hilbert space, and then explore this enlarged space in a numerically efficient way and find the optimal solution. In the context of clustering problem, the hope is that by exploring a much larger space, the algorithm is less likely to result in bad clustering. This paper is organized as follows. In Sec.~\ref{sec:method}, we describe our method in detail. In{\color{red}{ Sec.~\ref{sec:result}}}, we apply our method to several machine learning data sets, showing that our method could result in a high learning accuracy and is less likely to be trapped in local minimum. We conclude in Sec.\ref{sec:summary}.


\section{Method}\label{sec:method}

\subsection{Classical K-means algorithm}
Before we introduce our method, we first briefly review the standard classical K-means algorithm~\cite{Lloyd1982}. For $N$ input vectors labeled as $\vec{x}_n$ ($1\leq n\leq N$), the classical K-means algorithm works as follows: 1) initializing $k$ random vectors $\ccenter_m$ ($1\leq m\leq k$) as the initial centroids. 2) Dividing all the data into $k$ classes according their distances to the centroids, denoted as $\dC$. Concretely, for each data vector $\vec{x}_n$, one computes
\begin{align}
\dC_{n,m} = \vert \vec{x}_n - \ccenter_m\vert^2,
\end{align}
where $\vert \vec{v} \vert = \sqrt{\sum v_j^2}$ is the standard $2$-norm of a vector $\vec{v}$. Then $\vec{x}_n$ is assigned to the $j$-th cluster with $j = \argmax_m \dC_{n, m}$. After this step, the $N$ data vectors are divided into $k$ non-overlapping clusters. 3) Computing the new center of each class by minimizing its variances independently, namely by minimizing the following loss function
\begin{align}
f^C(\ccenter_m) = \sum_{n_j=1}^{N_j} \vert \vec{x}^j_{n_j} - \ccenter_m \vert^2 = \sum_{n_j=1}^{N_j} \dC_{n_j, m} ,
\end{align}
where $N_j$ denote the size of the $j$-th cluster, and $\vec{x}^j_{n_j}$ denotes the $n_j$-th vector inside the $j$-th cluster. By repeating steps 2) and 3), one could often quickly converge to a local minimum. In the prediction stage, given a new input vector $\vec{x}$, one simply computes each of the distance $\dC(\vec{x}, \ccenter_m)$ between $\vec{x}$ and the centroid $\ccenter_m$, and then $\vec{x}$ is assigned to the $j$-th cluster with $j = \argmax_m \dC(\vec{x}, \ccenter_m)$.

\subsection{Quantum K-means algorithm}
The above K-means algorithm could be straightforwardly converted into a quantum machine learning algorithm, which we describe as follows and will be referred to as the quantum K-means algorithm. In this case we assume that there is a list of input quantum states labeled as $\vert x_n\rangle$. The centers are denoted as $\qcenter{m}$ and are initialized using parametric quantum circuits~\cite{MitaraiFujii2018, LiuWu2020}
\begin{align}
\qcenter{m} = \circuit{m} \vert 0\rangle^{\otimes L},
\end{align}
where $\circuit{m}$ denotes the $m$-th parametric quantum circuit with an array of parameters denoted by $\vec{\theta}$. The distance between $\vert x_n\rangle$ and $\qcenter{m}$ can be defined as
\begin{align}
d^Q_{n, m} = 1 - \vert\langle x_n \qcenter{m} \vert^2,
\end{align}
where the second term on the right hand side of the above equation can be efficiently computed with a quantum computer using the SWAP test technique~\cite{BuhrmanWolf2001}. Similarly, the loss function for the $m$-th cluster can be defined as
\begin{align}
f^Q(\qcenter{m}) =\sum_{n_j=1}^{N_j} \left(1 - \vert\langle x_{n_j}^j \qcenter{m} \vert^2\right) = \sum_{n_j=1}^{N_j} \dQ_{n_j, m}.
\end{align}
We note that the gradient of $f^Q(\qcenter{m})$ can also be computed with a quantum computer using~\cite{MitaraiFujii2018}
\begin{align}
\frac{\partial f^Q(\qcenter{m})}{\partial \theta_i} = \frac{1}{2}f^Q(\qcenter{m}^-) - \frac{1}{2}f^Q(\qcenter{m}^+),
\end{align}
where $\qcenter{m}^{\pm}$ means to shift the $i$-th parameters $\theta_i$ in $\vec{\theta}$ by $\pm\frac{\pi}{2}$. Implementing SWAP test on current quantum computers is not an easy task since it requires the three-qubit TOFFILI gate. In this work we will not go any further on this quantum machine learning algorithm but will instead focus on the quantum inspired classical algorithm as follows.

\subsection{Quantum inspired K-means algorithm}
Now we will describe in detail our quantum inspired K-means algorithm. Similar to the quantum K-means algorithm, the inputs will be assumed to be a list of quantum states. However, here we also assume that each quantum state $\vert x_n\rangle$ could be parameterized using an MPS $X_n^{\vec{\sigma}}$
\begin{align}
X_n^{\vec{\sigma}} = \sum_{a_1, a_2, \dots, a_{L+1}}X^{\sigma_1}_{n, a_{1}, a_{2}} X^{\sigma_2}_{n, a_{2}, a_{3}}\dots X^{\sigma_L}_{n, a_{L}, a_{L+1}},
\end{align}
where $L$ is the number of qubits required by the corresponding quantum state $\vert x_n\rangle$. The size of an MPS is characterize by an integer $D$, referred to as the bond dimension, defined as
\begin{align}
D = \max_{1\leq l\leq L}\left(\dim(a_l)\right).
\end{align}
We note that for a vector $\vec{x}_l$ with $L$ elements $\left(x_{l, 1}, x_{l,2}, \dots, x_{l, L} \right)$, one could map it into a separable quantum state $\vert x_l\rangle$ such that each element $x_{n, l}$ is mapped into a single-qubit state $\cos(\frac{\pi}{2} x_{n, l})\vert 0\rangle + \sin(\frac{\pi}{2} x_{n, l})\vert 1\rangle$. This would correspond to a separable MPS $X_n^{\vec{\sigma}}$ of $D=1$ with elements
\begin{align}\label{eq:encoding}
X_{n,1,1}^{0} = \cos(\frac{\pi}{2}x_{n, l}), \quad X_{n,1,1}^{1} = \sin(\frac{\pi}{2}x_{n, l}).
\end{align}
The centroids are then assumed to be MPSs with a fixed bond dimension $D$, and written as $\mcenter{m}$
\begin{align}
\mcenter{m} = \sum_{b_1, b_2, \dots, b_{L+1}}M^{\sigma_1}_{m, b_{1}, b_{2}} M^{\sigma_2}_{m, b_{2}, b_{3}}\dots M^{\sigma_L}_{m, b_{L}, b_{L+1}}.
\end{align}
The distance between two MPSs is defined as
\begin{align}
\dMPS_{n, m} =\left(X_n^{\vec{\sigma}, \dagger} - \mcenterbra{m}\right) \left(X_n^{\vec{\sigma}} - \mcenter{m} \right),
\end{align}
and similarly, the loss function for the $m$-th cluster can be defined as
\begin{align}
f^M(\mcenter{m}) = \sum_{n_j=1}^{N_j} \dMPS_{n_j, m},
\end{align}
with the normalization condition $\mcenterbra{m}\mcenter{m}=1$. The above cost function can be minimized by setting the gradient against each tensor $M_{m, b_l, b_{l+1}}^{\sigma_l}$ to $0$, namely
\begin{align}\label{eq:partialgrad}
\frac{\partial f^M(\mcenter{m})}{\partial M^{\sigma_l}_{m, b_{l}, b_{l+1}}} = 0.
\end{align}
The above equation can be simply reduced to a set of algebraic equations. To see this, we first define
\begin{align}
L_{b_{l}}^{b_{l}'} &= \sum_{b_{l-1}, b_{l-1}', \sigma_{l-1} }L_{b_{l-1}}^{b_{l-1}'} M^{\sigma_{l-1}}_{b_{l-1}, b_{l}} M^{\sigma_{l-1}}_{b_{l-1}', b_{l}'}; \\
R_{b_{l+1}}^{b_{l+1}'} &= \sum_{b_{l+2}, b_{l+2}', \sigma_{l+1} }R_{b_{l+2}}^{b_{l+2}'} M^{\sigma_{l+1}}_{b_{l+1}, b_{l+2}} M^{\sigma_{l+1}}_{b_{l+1}', b_{l+2}'}; \\
A_{b_{l}}^{a_{l}} &= \sum_{b_{l-1}, a_{l-1}, \sigma_{l-1} }A_{b_{l-1}}^{a_{l-1}} M^{\sigma_{l-1}}_{b_{l-1}, b_{l}} X^{\sigma_{l-1}}_{a_{l-1}, a_{l}}; \\
B_{b_{l+1}}^{a_{l+1}} &= \sum_{b_{l+2}, a_{l+2}, \sigma_{l+1} }R_{b_{l+2}}^{a_{l+2}} M^{\sigma_{l+1}}_{b_{l+1}, b_{l+2}} M^{\sigma_{l+1}}_{a_{l+1}, a_{l+2}},
\end{align}
with $L_{b_{1}}^{b_{1}'} = R_{b_{L+1}}^{b_{L+1}'} = A_{b_{1}}^{a_{1}} = B_{b_{L+1}}^{a_{L+1}} = 1$. With these equations, Eq.~\ref{eq:partialgrad} can be written as
\begin{align}
L_{b_{l}}^{b_{l}'} M^{\sigma_l}_{b_{l}, b_{l+1}} R_{b_{l+1}}^{b_{l+1}'} = \sum_{i=1}^{n} A_{b_{l}}^{a_{l}} X^{\sigma_l}_{i, a_l, a_{l+1}} B_{b_{l+1}}^{a_{l+1}},
\end{align}
where we have neglected the label $m$ for the cluster since each cluster can be minimized independently. Interestingly, if we keep $\mcenter{m}$ in a mixed-canonical form~\cite{Schollwock2011}, then $L_{b_{l}, b_{l}'} = \delta_{b_{l}, b_{l}'}$ and $R_{b_{l+1}, b_{l+1}'} = \delta_{b_{l+1}, b_{l+1}'}$, such that the above equation reduces to
\begin{align}
M^{\sigma_l}_{b_{l}, b_{l+1}} = \sum_{n=1}^{N} A_{b_{l}}^{a_{l}} X^{\sigma_l}_{n, a_l, a_{l+1}} B_{b_{l+1}}^{a_{l+1}}.
\end{align}


\section{Applications}\label{sec:result}

\begin{table}[!htb]
\caption{Datasets used for comparison between classical K-means algorithm and our quantum inspired algorithm.}
\label{tab:tab1}
\centering
\label{tab1}
    \begin{tabular}{|c|c|c|c|}
        \hline
        Name & Points & Dimensions & Class\\
        \hline
        Breast & 683 & 9 & 2 \\
        \hline
        Ionosphere & 351 & 34 & 2\\
        \hline
        Wine & 178 & 13 & 3\\
        \hline
        Yeast & 1484 & 8 & 10\\
        \hline
        E-coli & 336 & 7 & 8\\
        \hline
    \end{tabular}
\end{table}

To demonstrate our method, we apply it to several commonly used machine learning datasets for clustering tasks as shown in TABLE~\ref{tab:tab1}. For each dataset, we randomly pick $80\%$ of the data as the training data, and the remaining $20\%$ as the testing data. To ensure a fair comparison, we will always use the same splitting of training and testing data for the classical K-means algorithm and our quantum inspired K-means algorithm. Since the K-means algorithm is based on the distance measurement, if the difference between variables of different dimensions is too large, it may cause a small number of variables to exert an excessively high influence on the whole cost, thus eliminating the effect of the rest variables. To avoid this effect, we divide each dimension of the data by the its largest value in the training set, namely
\begin{align}
x_{n, j} \leftarrow \frac{x_{n, j}}{A^{\max}_{j}},
\end{align}
with $A^{\max}_{j} = \max_{1\leq n\leq N_{train}} x_{n, j}$, where $N_{train}$ is the total number of training data. As a result, we have $0\leq x_{n, j}\leq 1$ for all $1\leq n\leq N_{train}$ and $1\leq j\leq L$. In the prediction stage, we also divide each dimension of the testing data by $A^{\max}_{j}$. We note that this procedure may produce elements larger than $1$ in the testing set, which may not be well distinguished by our quantum inspired algorithm since our encoding in Eq.~\ref{eq:encoding} is a periodic function of period $1$. However, our numerical results show that we could still get high precision results using our method despite those elements.

In the following, we show the comparison between our quantum inspired K-means algorithm with the classical K-means algorithm in terms learning accuracy and the likelihood of getting trapped in local minima. In TABLE.~\ref{tab:tab2}, we show the accuracies for different datasets, which is defined as the number of the correctly predicted data divided by the total number of testing data. Since it is possible for both algorithms to be trapped in local minima, we run the same simulation for each algorithm for $50$ times with randomly initialized centroids and pick the one with the highest accuracy. For our quantum inspired K-means algorithm, we also tune the bond dimension $D$ to be $8$ and $15$ to see the effect of different bond dimensions. We can see from TABLE.~\ref{tab:tab2} that with $D=8$, our quantum inspired algorithm already performs as good as or better than the classical algorithm for all the cases. And for $D=15$, we get even higher accuracies.

\begin{table}[!htb]
\caption{Comparison of accuracies on the testing data between the classical K-means algorithm and the quantum inspired K-means algorithm. For the E-coil dataset, the data for the case $D=15$ is missing since the dimension of each data is too small.}
\label{tab:tab2}
\centering
\label{tab1}
    \begin{tabular}{|c|c|c|c|}
        \hline
        \multirow{3}{*}{} & \multirow{3}{*}{Classic k-means}  & \multicolumn{2}{|c|}{Quantum inspired k-means}\\
        \cline{3-4}

         & & D = 8 & D = 15\\


        \hline
        Breast & 99.27$\%$ & 100$\%$ & 100$\%$ \\
        \hline
        Ionosphere & 60.56$\%$ & 60.56$\%$ & 63.38$\%$\\
        \hline
        Wine & 100$\%$ & 100$\%$ & 100$\%$\\
        \hline
        Yeast & 45.45$\%$ & 47.47$\%$ & 48.48$\%$\\
        \hline
        E-coli & 85.29$\%$& 89.71$\%$ & none \\
        \hline
    \end{tabular}
\end{table}

\begin{figure}
	\centering
	\includegraphics[width=3.5in]{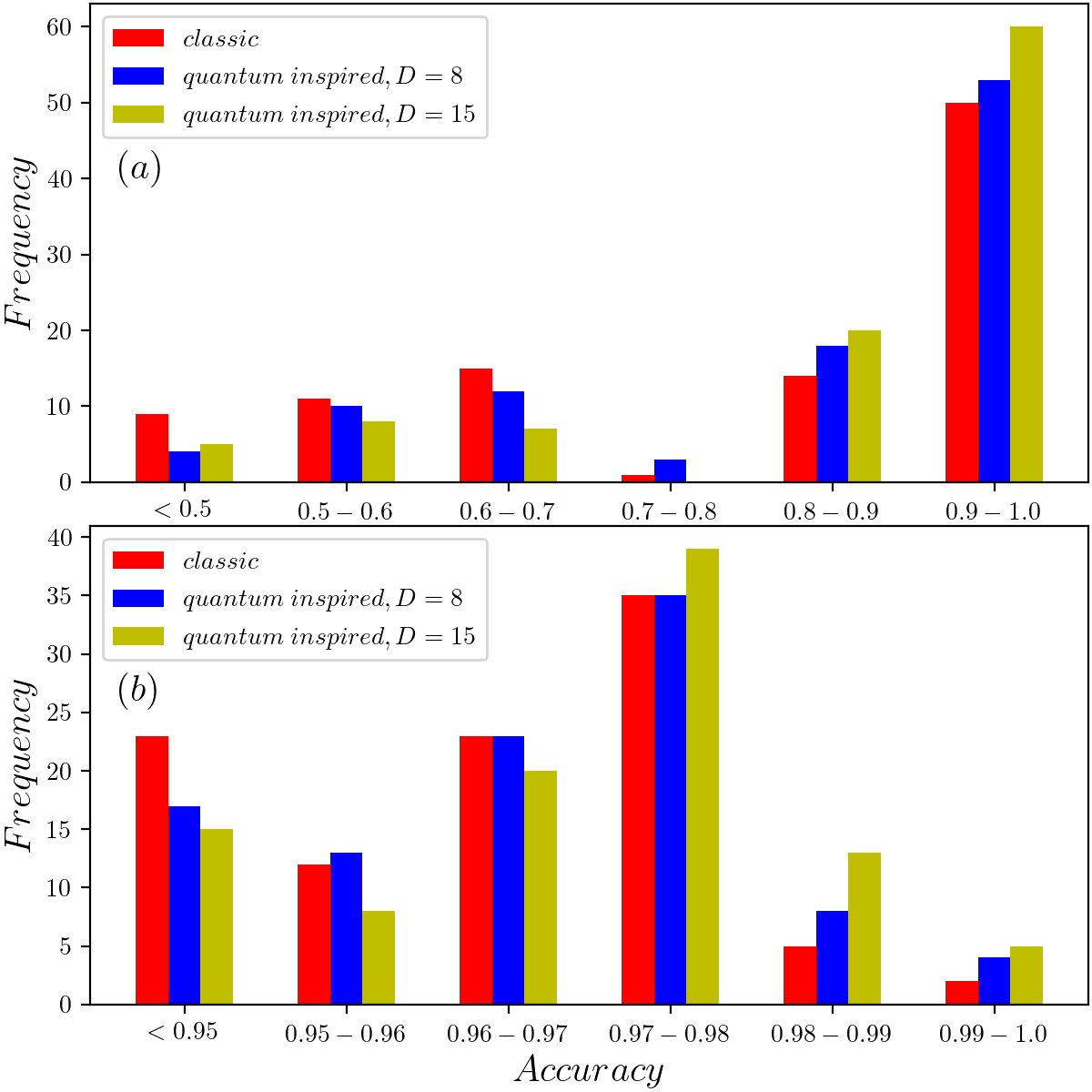}
	\caption{Frequency as a function of the accuracy for the Wine dataset (a), and the Breast dataset (b). The red, blue, yellow columns correspond to classical K-means, quantum inspired K-means algorithm with $D=8$ and $D=15$ respectively. The y-axis is the number of instances out of $100$ cases with an accuracy in between the interval indicated by the x-axis.}
	\label{fig:fig2} 
\end{figure}

A well-known drawback of the classical K-means algorithm is that it could easily be trapped in local minima. Here we also check the likelihood of our algorithm to be trapped in local minima. For this purpose, we take the Wine and Breast dataset as examples and we run the same simulation for $100$ randomly initialized centroids for each algorithm. We show the distribution of the number of cases against different accuracies as a histogram in Fig.~\ref{fig:fig2}. We can see that for our quantum inspired K-means algorithm, there is a higher probability to get a higher accuracy. For example, for the Wine dataset, there are $50$ cases out of $100$ with accuracies beyond $0.9$ for the classical K-means algorithm, while for quantum inspired K-means algorithm with a bond dimension $D=8$ there are $53$ cases, and for $D=15$ there are $60$ cases.

Another way to access the quality of the learned centers is to compute the Euclidean distances between the centroids in both cases. It is already shown in~\cite{SunSu2020} that the classical to quantum encoding in Eq.\ref{eq:encoding} could result in better separated input data, thus when appropriate learning algorithms are used, it is more likely to get higher prediction accuracy. In the classical case, The Euclidean distance between two different centers $\ccenter_i$ and $\ccenter_j$ is
\begin{align}
D_{i,j}= |\ccenter_i - \ccenter_j |,
\end{align}
where $|\vec{v}|$ means the $2$-norm of the vector $\vec{v}$. Similarly, in the quantum inspired case, the distance between two different centers $\mcenter{i}$ and $\mcenter{j}$ is

\begin{align}
D^{\vec{\sigma}}_{i, j} = |\langle\mcenter{i} | \mcenter{j}\rangle|^2.
\end{align}

\begin{figure}
	\centering
	\includegraphics[width=3.5in]{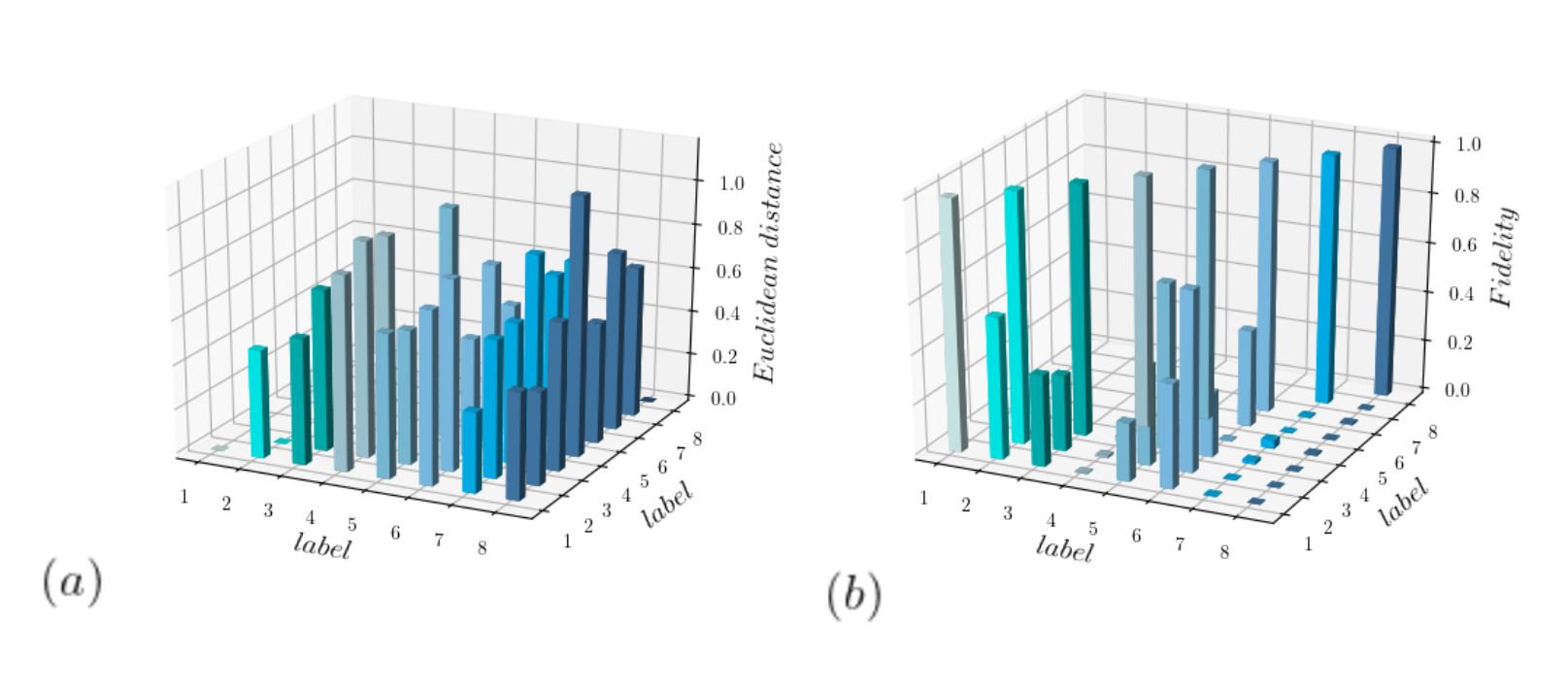}
	\caption{(a) Distances between learned centroids for the classical K-means algorithm. (b) Distances between learned centroids for the quantum inspired K-means algorithm. Here the E-coil dataset is used.}
	\label{fig:fig3} 
\end{figure}

Since $\mcenter{i}$ lives in a much larger space than $\ccenter_i$, it is possible that the centroids are more likely orthogonal to each other in the quantum inspired case, that is, the $k\times k$ matrix formed by $D^{\vec{\sigma}}_{i, j}$ is closer to a diagonal matrix. As a result, it is easier to label a new data in the correct category. To show this clearly, we directly show the distance matrices formed by $D_{i,j}$ and $D^{\vec{\sigma}}_{i, j}$ based on the optimized centroids for the E-coil dataset in Fig.\ref{fig:fig3}. We can see clearly that in the classical case the centroids have large overlaps between one another while in the quantum inspired case, the centers are almost orthogonal to each other.

\begin{figure}
	\centering
	\includegraphics[width=3.5in]{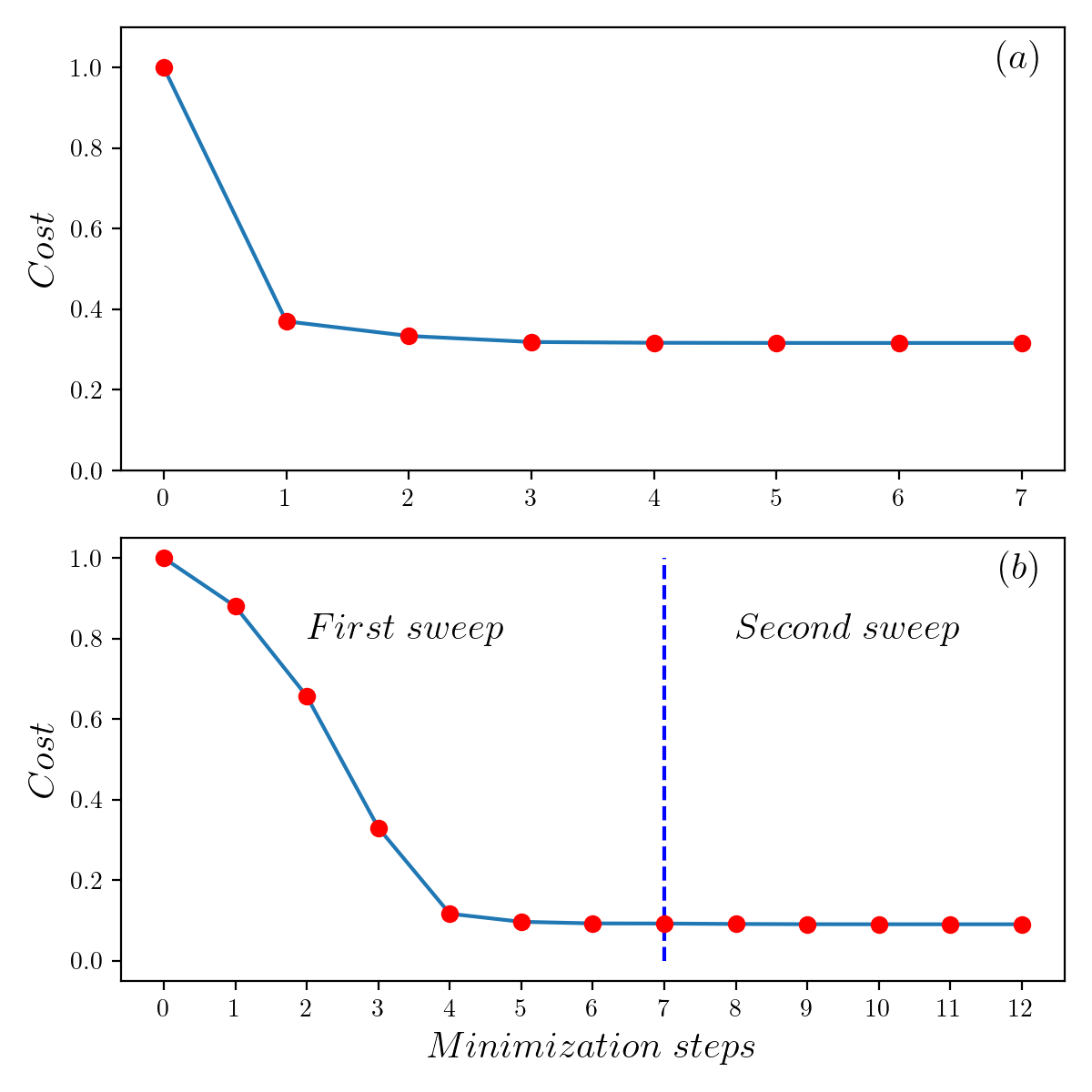}
	\caption{(a) Loss value as a function of K-means iterations for the classical K-means algorithm. (b) Loss value as a function of K-means iterations. In each K-means iteration we use $1$ variational-MPS sweep which contains $2(L-1)$ local minimizations steps.}
	\label{fig:fig4} 
\end{figure}

Lastly, we compare the convergence rate towards minima for different algorithms. In Fig.\ref{fig:fig4}, we show the loss values as a function of the minimization steps. In the classical case, the loss values are computed after each K-means iteration which is shown in Fig.\ref{fig:fig4}(a). Each K-means iteration is counted as one minimization step in this case. In the quantum inspired case, in each K-means iteration, we use a single variation MPS sweep in which the local on-site energy is first minimized from the left boundary to the right boundary and then back from the right boundary to the left boundary~\cite{Schollwock2011}. Each local minimization is counted as a minimization step in the quantum inspired case. In Fig.\ref{fig:fig4}(b), we show the loss values in this case. We have taken the initial loss value as $1$ in Fig.~\ref{fig:fig4} and the rest loss values are renormalized against the initial value. We can see that the classical K-means algorithm takes $7$ minimization steps ($7$ K-means iterations) to reach a final loss value $0.316$, while the quantum inspired K-means algorithm uses $12$ minimization steps ($2$ K-means iterations) to reach a final loss value $0.091$.



\section{Conclusion}\label{sec:summary}
In this paper, we propose a quantum inspired K-means clustering algorithm based on matrix product states. By mapping the input data into a much larger quantum Hilbert space and clustering data in the enlarged space, we show that the learning algorithm could result in a much higher precision. We demonstrate our algorithm by applying it to several commonly used machine learning datasets. Our results show that our algorithm is advantageous to the classical K-means algorithm in that 1) our algorithm could reach a higher prediction accuracy and 2) our algorithm is less likely to be trapped in local minima. We show that compared to the classical K-means algorithm, the learned centroids with our algorithm are better separately, which could be a reason why it could make more precise predictions. We also show the loss values as a function of the number of minimizations steps to help to better visualize the convergence in both algorithms.

To this end, we point out that there are at least two directions which could be inspired from this work: 1) the classical K-means algorithm is well studied and there exists various techniques to improve the learning accuracy on top of the standard K-means algorithm. In future works those techniques could also be adapted into our quantum inspired algorithm to further increase the learning accuracy; 2) As discussed in Sec.\ref{sec:method}, our quantum inspired algorithm has a one-to-one correspondence with a pure quantum machine learning algorithm which could be readily be executed on a quantum computer, with the most significant difference that in the quantum inspired algorithm the quantum state is represented using matrix product states which could be efficiently stored on a classical computer. In the near future it would be interesting to carry out the quantum K-means algorithm on a quantum computer with applications to real world clustering problems. Other than that, one could also explore the two-site variant of the quantum inspired algorithm such that the bond dimension $D$ could be dynamically adjusted. Moreover, in this work we use variational-MPS sweeps where we minimized the local energy on each site iteratively, it could also be interesting to compare this approach with the gradient-based minimization methods.




\section{Acknowledgement}
 Y. Shang thanks the support of National Key Research and Development Program of China under grant 2016YFB1000902, National Research Foundation of China (Grant No. 61872352, 61472412), and Program for Creative Research Group of National Natural Science Foundation of China (Grant No. 61621003); C. Guo. acknowledges support from National Natural Science Foundation of China under Grants No. 11805279.

\bibliographystyle{apsrev4-1}
\bibliography{refs}

\end{document}